\documentclass[10pt]{iopart}
\pdfoutput=1
\usepackage{graphicx,color}
\begin{document}

\title{The critical catastrophe revisited}

\author{Cl\'elia de Mulatier}
\address{CNRS - Universit\'e Paris-Sud, LPTMS, UMR8626, 91405 Orsay Cedex, France}
\address{CEA/Saclay, DEN/DANS/DM2S/SERMA/LTSD, 91191 Gif-sur-Yvette, France}
\author{Eric Dumonteil}
\address{CEA/Saclay, DEN/DANS/DM2S/SERMA/LTSD, 91191 Gif-sur-Yvette, France}
\author{Alberto Rosso}
\address{CNRS - Universit\'e Paris-Sud, LPTMS, UMR8626, 91405 Orsay Cedex, France}
\author{Andrea Zoia}
\address{CEA/Saclay, DEN/DANS/DM2S/SERMA/LTSD, 91191 Gif-sur-Yvette, France}
\ead{andrea.zoia@cea.fr}

\vspace{10pt}

\begin{indented}
\item[]June 2015
\end{indented}

\begin{abstract}
The neutron population in a prototype model of nuclear reactor can be described in terms of a collection of particles confined in a box and undergoing three key random mechanisms: diffusion, reproduction due to fissions, and death due to absorption events. When the reactor is operated at the critical point, and fissions are exactly compensated by absorptions, the whole neutron population might in principle go to extinction because of the wild fluctuations induced by births and deaths. This phenomenon, which has been named critical catastrophe, is nonetheless never observed in practice: feedback mechanisms acting on the total population, such as human intervention, have a stabilizing effect. In this work, we revisit the critical catastrophe by investigating the spatial behaviour of the fluctuations in a confined geometry. When the system is free to evolve, the neutrons may display a wild patchiness (clustering). On the contrary, imposing a population control on the total population acts also against the local fluctuations, and may thus inhibit the spatial clustering. The effectiveness of population control in quenching spatial fluctuations will be shown to depend on the competition between the mixing time of the neutrons (i.e., the average time taken for a particle to explore the finite viable space) and the extinction time.
\end{abstract}

%
\vspace{2pc}
%
\submitto{\JSTAT}
%
%
%

\section{Introduction}
\label{intro}

Many physical and biological systems can be represented in terms of a collection of individuals governed by the competition of the two basic random mechanisms of birth and death. Examples are widespread and encompass neutron multiplication~\cite{pazsit, williams, harris}, nuclear collision cascades~\cite{harris, barucha, athreya}, epidemics and ecology~\cite{bailey, jagers, murray, zhang, meyer}, bacterial growth~\cite{golding, houchmandzadeh_prl}, and genetics~\cite{lawson, bertoin, sawyer}. Neglecting particle-particle correlations and non-linear effects, the evolution of such systems can be effectively explained by the Galton-Watson model~\cite{harris}. When the death rate is larger than the birth rate, the system is said to be sub-critical: the population size decreases on average, and the ultimate fate is extinction. This occurs for instance for nuclear collision cascades, where charged particles are progressively scattered and absorbed by the medium~\cite{harris, barucha}. When on the contrary the birth rate is larger than the death rate, such as for bacteria reproducing on a Petri dish~\cite{houchmandzadeh_prl}, the system is said to be super-critical. In this case, the population size grows on average. However, because of fluctuations on the number of individuals in the population, a non-trivial finite extinction probability exists for the whole system~\cite{harris}. A super-critical regime is typically found also during the early stages of an epidemic (the so-called `outbreak' phase), where a fast growth of the infected population is observed, until non-linear effects due to the depletion of susceptible individuals ultimately slow down the epidemic~\cite{pnas}.

In the intermediate regime, the population stays constant on average, and the system is said to be exactly critical. A prominent example of a system operating at the critical point is provided by the self-sustaining chains of neutrons in nuclear reactors~\cite{pazsit, williams, sanchez}. Avalanches and self-organized criticality are other examples of system operating at or close to the critical point. At the critical or nearly critical regime, fluctuations due to birth and death may become particularly strong~\cite{harris}. Assuming without loss of generality that the birth and death rates are Poissonian, from the Galton-Watson theory it is known that at criticality the total number $N(t)$ of particles in the system stays constant on average, i.e., $\langle N \rangle = N_0$, whereas the variance grows in time, i.e., $\textit{Var}[N] = \langle N^2\rangle - \langle N\rangle^2 \propto \lambda N_0 t$, where $\lambda$ is the birth/death rate~\cite{harris}. This immediately implies that the typical fluctuations of the population size, say $\sqrt{\textit{Var}[N]}$, will become comparable to the average population size $N_0$ over a time $\sim N_0/\lambda$. Hence, a critical system will have a characteristic extinction time of the order of $\tau_E \simeq N_0/\lambda$~\cite{harris}. This quite extreme behaviour has been observed for instance in avalanche dynamics. In the context of neutron multiplication, the shut-down of a reactor operated in the critical regime due to the extinction of the fission chains has been theoretically investigated~\cite{pazsit, williams, sanchez, mechitoua} and goes under the name of critical catastrophe~\cite{williams}. However, such observation is in open contradiction with the experience: the behavior of nuclear reactors at the critical point is actually stable. This apparent paradox has been explained by pointing out that including feedback mechanisms (representing for instance human intervention) in the Galton-Watson model induces a stabilizing effect acting against the total population fluctuations~\cite{williams}.

\begin{figure}
  \includegraphics[width=8cm]{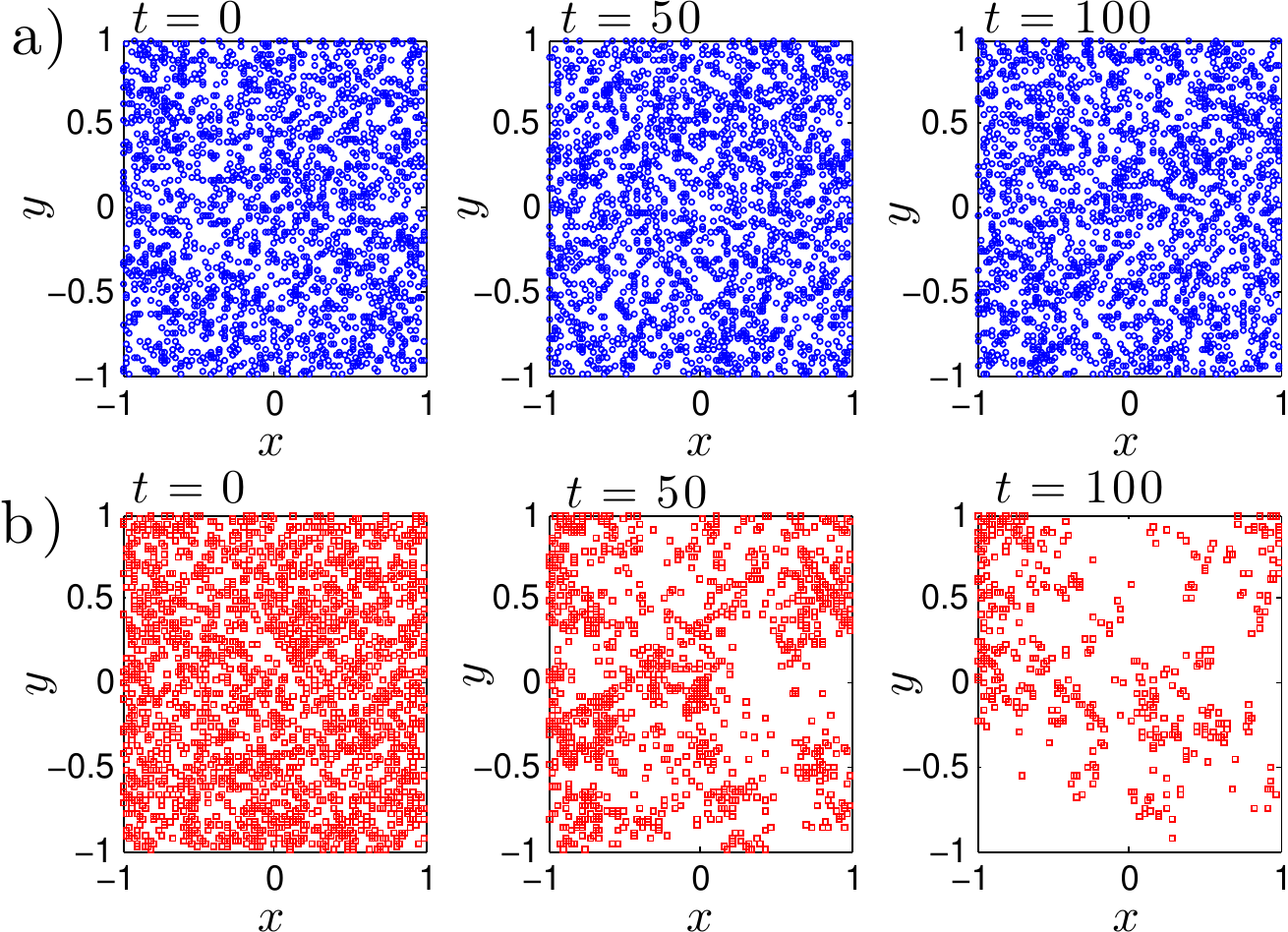}
\caption{Monte Carlo simulation of the evolution of a collection of particles in a two-dimensional box. The particles are prepared at $t=0$ on a uniform spatial distribution. In case $a)$, particles follow regular Brownian motions: as time increases, positions are shuffled by diffusion, but the spatial distribution of the particles stays uniform. In case $b)$, particles follow a branching Brownian motion with equal birth and death rates: as time increases, the population undergoes large fluctuations, and the particle density displays a wild patchiness. Eventually, the entire population goes to extinction.}
\label{fig1}
\end{figure}

In most of the models cited above, individuals also interact with the surrounding environment and are typically subject to random displacements~\cite{williams, spatial_eco, zhang, meyer}. The interplay between the fluctuations stemming from birth-death events and those stemming from diffusion will thus subtly affect the spatial distribution of the particles in such systems~\cite{legall, derrida, derrida_barrier, berestycki_three, majumdar}. In particular, it has been shown that at and close to the critical point a collection of such individuals, although spatially uniform at the initial time, may eventually display a wild patchiness (see Fig.~\ref{fig1}), with particles closely packed together and empty spaces nearby~\cite{zhang, meyer, young, houchmandzadeh_prl}. Spatial clustering phenomena have been first identified in connection with mathematical models of ecological communities~\cite{cox, dawson}, and since then have been thoroughly investigated for both infinite and finite collections of individuals~\cite{zhang, meyer, young, houchmandzadeh_pre_2002, houchmandzadeh_pre_2009, zoia_pre_clustering}. Non-uniform neutron densities in the reactor might lead to local peaks in the deposited energy (hot spots) and represent thus a most unwanted event with respect to the safe operation of nuclear power plants~\cite{dumonteil_ane}. In view of the findings concerning the impact of a feedback on the global neutron population, we might wonder whether imposing a global feedback on the total neutron population affects also the local spatial behaviour of the particles.

In this work, we will revisit the critical catastrophe of neutron chains in a prototype model of a nuclear reactor, with special emphasis on the spatial distribution of neutrons in confined geometries. We will show that actually a global feedback has a stabilizing effect also on the local particle density, and may thus inhibit clustering. In particular, the effectiveness of population control in quenching spatial fluctuations will be shown to depend on the competition between the mixing time of the neutrons (i.e., the average time taken for a particle to explore the finite viable space) and the extinction time. This paper is organized as follows: in Sec.~\ref{prototype_model} we will sketch the statistical model of the neutron population in a nuclear reactor. Then, in Sec.~\ref{analysis_correlations} we will illustrate the main findings concerning the pair correlation functions for a collection of particles in a confined box with and without population control. The detailed derivation of the pair correlation functions will be discussed in Sec.~\ref{analysis_correlations}, and conclusions will be finally drawn in Sec.~\ref{conclusions}. Calculations for one-dimensional systems will be worked out in the Appendix.

\section{A prototype model of nuclear reactor}
\label{prototype_model}

\begin{figure}
  \includegraphics[width=8cm]{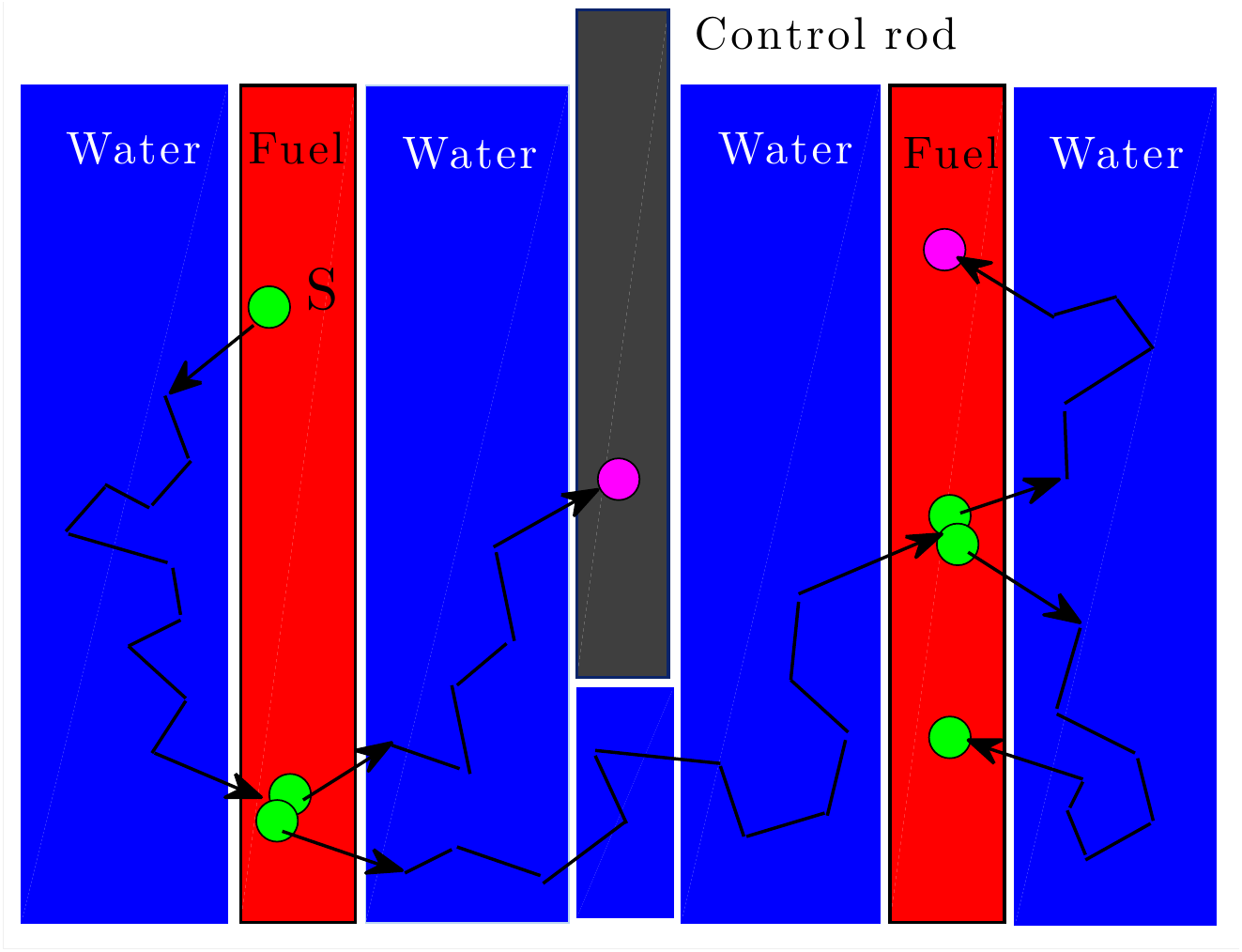}
\caption{Simplified scheme of neutron propagation within a nuclear reactor. A fission chain begins with a source neutron (marked with $S$ in the figure) born from a fission event in the fuel. The neutron diffuses in the water and may eventually come back to a fuel element. Then, it can either be absorbed (these events are marked by magenta circles), in which case the trajectory terminates; or it can give rise to a new fission, upon which additional neutrons are set free (these events are marked in green), and the fission chain is kept alive. The system is operated at the critical point when the average net number of neutrons produced at fission is exactly compensated by the losses by absorptions. A control rod can absorb the excess neutrons so as to adjust the total population and enforce the critical regime.}
\label{fig2}
\end{figure}

A nuclear reactor is a device conceived to extract energy from the fission chains induced by neutrons~\cite{bell}. To fix the ideas, here we will focus on the widely used light-water reactors. The nuclear fuel is made of uranium, arranged in a regular lattice and plunged in light water. A fission chain begins with a neutron emitted at high energy from a fission event (see Fig.~\ref{fig2}). The neutron enters the surrounding water, slows down towards thermal equilibrium, and then starts diffusing. If the neutron eventually re-enters the fuel, it may $i)$ be absorbed on the $^{238}U$ isotope of uranium, in which case the chain is terminated; or $ii)$ give rise to a new fission event by colliding with the $^{235}U$ fissile isotope, whereupon a random number of high-energy neutrons are emitted. The water surrounding the fuel lattice acts as a reflector and prevents the neutrons from escaping from the core. A number of control rods are inserted into the core, with the aim of absorbing the excess neutrons and keep the population constant (this ensures a constant power output). When the neutron population grows, the control rods are inserted more deeply into the core, slowing down the chain reaction. On the contrary, when the population decreases, the control rods are raised, accelerating the chain reaction.

Nuclear reactors are complex devices: their energy- and spatial-dependent behaviour can be fully assessed only by resorting to large-scale numerical simulations including a realistic description of the heterogeneous geometry~\cite{dumonteil_ane}. However, for the purposes of this work we will introduce a simplified prototype model of a nuclear reactor that yet retains all the key ingredients of a real system.

We will assume that the reactor can be represented as a collection of $N$ neutrons undergoing diffusion, reproduction and absorption within an homogeneous $d$-dimensional box of finite volume $V$, with reflecting (mass-preserving) boundaries. It is reasonable to require that the initial neutron population has a uniform spatial distribution. The stochastic paths of neutrons are known to follow position- and velocity-dependent exponential flights~\cite{williams, exp_flights, zoia1}. For our model, we approximate these paths by regular $d$-dimensional Brownian diffusion with a constant diffusion coefficient $D$. The diffusing walker undergoes a birth-death event at rate $\lambda$: the neutron disappears and is replaced by a random number $m$ of descendants ($m=0$ accounting for absorption), distributed according to a law $q_m$ with average $\nu_1 = \sum_m m q_m$. In order for the reactor to be exactly critical, we must have $\nu_1=1$.

\begin{figure}
  \includegraphics[width=11cm]{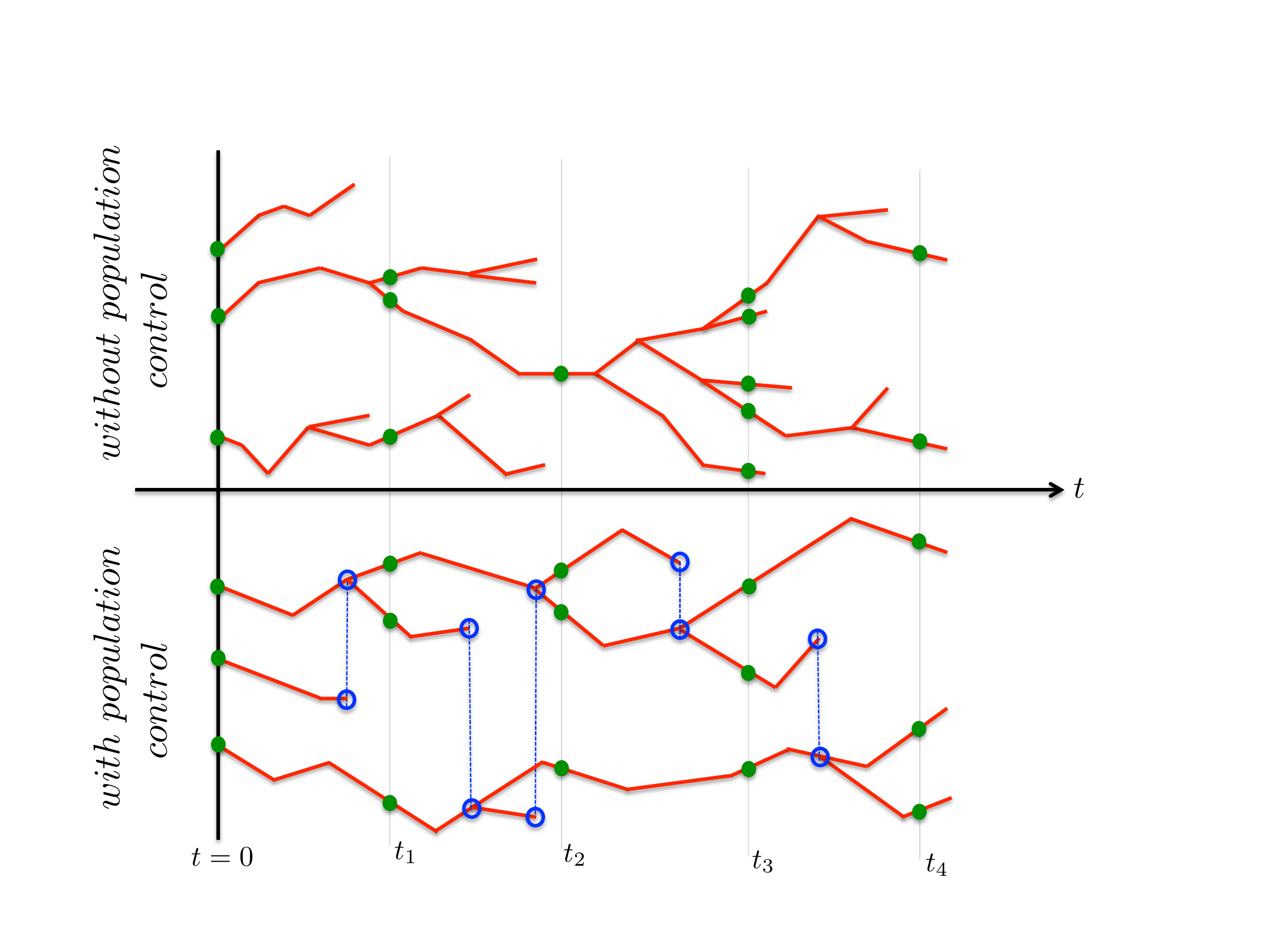}
\caption{The evolution of a collection of branching random walks with binary fission and $\nu_1=1$. At time $t=0$, $N=3$ particles are present, and the system is observed at successive times $t_1 < t_2 < t_3 < t_4$. Top. When population control is not enforced, the number of particles present at the observation times fluctuates because births and deaths occur at random instants. Bottom. When population control is enforced, at each fission event a neutron is randomly chosen and removed, which exactly preserves $N$: at observation times, the total population is constant.}
\label{fig3}
\end{figure}

Clustering phenomena have been mostly analysed either in the thermodynamic limit ($V \to \infty$ and $N \to \infty$, with finite $N/V$~\cite{cox, houchmandzadeh_pre_2009}) or in unbounded domains with finite $N$~\cite{zhang, meyer}. A realistic description of actual physical systems demands however that the effects due to the finiteness of the viable volume $V$ be explicitly taken into account. In a previous work, we have shown that a neutron population within a finite-size reactor at the critical regime will ultimately undergo wild spatial fluctuations over the entire volume~\cite{zoia_pre_clustering}. In this paper, we include the effects of population control by imposing that the total number $N$ of neutrons in $V$ is preserved, and investigate the consequences of such constraint on spatial fluctuations. The simplest way to enforce a constant $N$ is to correlate reproduction and absorption events~\cite{zhang, meyer}: at each fission, a neutron disappears and is replaced by a random number $m \ge 1$ of descendants, and $m-1$ other neutrons are simultaneously removed from the collection in order to ensure the conservation of total population (see Fig.~\ref{fig3}). This mechanism has been first introduced in the theoretical ecology (with binary branching $q_m = \delta_{m,2}$~\cite{zhang, meyer}), where similar large-scale constraints such as limited food resources have been shown to quench the wild fluctuations in the number of individuals that are expected for an unconstrained community. Similar effects have been also considered in the context of cellular growth under the effects of chemotaxis~\cite{golestanian}.

\section{Physical observables and main findings}
\label{analysis_correlations}

Let us denote by $n({{\mathbf x}_i},t)$ the instantaneous density of neutrons located at ${{\mathbf x}_i}$ at time $t$. For an exactly critical system ($\nu_1=1$), the average neutron density at a point ${\mathbf x}_i$ reads
\begin{equation}
\langle n({{\mathbf x}_i},t)\rangle =N \rho({\mathbf x}_i,t),
\end{equation}
where we have set
\begin{equation}
\rho({\mathbf x}_i,t) = \int d{\mathbf x}_0 Q({\mathbf x}_0) {\cal G}({\mathbf x}_i, {\mathbf x}_0, t).
\end{equation}
Here $Q$ is the spatial probability distribution function of the neutrons at time $t=0$, and the propagator ${\cal G}({\mathbf x}, {\mathbf x}_0,t)$ times $d{\mathbf x}$ gives the probability to find a regular Brownian particle inside the volume element $({\mathbf x}; {\mathbf x}+d{\mathbf x})$ at time $t$, knowing that the particle was at ${\mathbf x}_0$ at time $t = 0$. The propagator satisfies the diffusion equation
\begin{equation}
\frac{\partial}{\partial t} {\cal G}({\mathbf x}, {\mathbf x}_0,t) =D \nabla^2_{{\mathbf x}_0} {\cal G}({\mathbf x}, {\mathbf x}_0,t),
\label{green_eq}
\end{equation}
with the appropriate boundary conditions. Note that at criticality the average neutron density becomes indistinguishable from that of $N$ regular Brownian particles. This result holds true independently of whether population control is applied. For an initial uniform source of particles $Q=1/V$, it immediately follows that for a collection of $N$ critical branching Brownian motions we simply have a uniform average density $\langle n({{\mathbf x}_i},t)\rangle = N/V$, at any time.

In order to probe the spatial inhomogeneities of the neutron population due to clustering, we must therefore go beyond the average behaviour. A fundamental tool is provided by the two-point (or pair) correlation function $h({{\mathbf x}_i},{{\mathbf x}_j},t)$ between positions ${{\mathbf x}_i}$ and ${{\mathbf x}_j}$, namely, the average density of pairs with the former particle in ${{\mathbf x}_i}$ and the latter in ${{\mathbf x}_j}$. This quantity is proportional to the joint probability density for ${{\mathbf x}_i}$ and ${{\mathbf x}_j}$~\cite{meyer}. For $N$ independent random walkers (in absence of branching and death) we simply have
\begin{equation}
h_{\textit{i}}({{\mathbf x}_i},{{\mathbf x}_j},t) = N (N-1) \rho({{\mathbf x}_i},t) \rho({{\mathbf x}_j},t).
\end{equation}
In particular, if the particles are uniformly distributed, $h_{\textit{i}}({\mathbf x}_i, {\mathbf x}_j, t) =N(N-1)/V^2$. More generally, the spatial shape of $h({{\mathbf x}_i},{{\mathbf x}_j},t)$ conveys information on the correlation range, whereas its amplitude is proportional to the correlation strength. A flat shape implies that the correlations have the same intensity everywhere; on the contrary, the presence of a peak at ${{\mathbf x}_i} \simeq {{\mathbf x}_j}$ reflects the increased probability of finding particles lying at short distances, which is the signature of spatial clustering~\cite{zhang, meyer, young, houchmandzadeh_pre_2009, houchmandzadeh_prl}. A closely related quantity is the average square distance between particles, i.e.,
\begin{equation}
\langle r^2 \rangle(t) = \frac{\int d{{\mathbf x}_i} \int d{{\mathbf x}_j} |{{\mathbf x}_i} - {{\mathbf x}_j}|^2 h({{\mathbf x}_i},{{\mathbf x}_j},t) }{\int d{{\mathbf x}_i} d{{\mathbf x}_i}  h({{\mathbf x}_i},{{\mathbf x}_j},t) },
\label{x2_definition}
\end{equation}
which is to be compared to the ideal average square distance of an uncorrelated population uniformly distributed in the available volume, namely,
\begin{equation}
\langle r^2 \rangle_{ \textit{id}} = \frac{1}{V^2} \int d{{\mathbf x}_i} \int d{{\mathbf x}_j} |{{\mathbf x}_i} - {{\mathbf x}_j}|^2 = \frac{d}{6} V^{\frac{2}{d}},
\end{equation}
where $d$ denotes the spatial dimension. Deviations of $\langle r^2 \rangle(t)$ from the reference value $\langle r^2 \rangle_{ \textit{id}}$ allow detecting spatial effects due to clustering~\cite{zhang, meyer}.

Analysis of the model detailed above shows that the population dynamics is governed by two distinct time scales: a mixing time $\tau_D \propto V^{2/d} / D$ and an extinction time $\tau_E \propto N/\lambda$. The quantity $\tau_D$ physically represents the time over which a particle has explored the finite viable volume $V$ by diffusion. Observe that the emergence of the time scale $\tau_D$ is a distinct feature of confined geometries having a finite spatial size: for unbounded domains, $\tau_D \to \infty$. The quantity $\tau_E$ has a different meaning according to whether population control is imposed~\cite{zhang, meyer}. For a free system, $\tau_E$ represents the time over which the fluctuations due to births and deaths lead to the extinction of the whole population. For a constrained system with constant $N$, $\tau_E$ represents the time over which the system has undergone a population renewal, and all the individuals descend from a single common ancestor. When the concentration $N/V$ of individuals in the population is large (and the system is spatially bounded), it is reasonable to assume that $\tau_E > \tau_D$. Intuitively, the precise shape of the pair correlation function must then depend on the subtle interplay of $\tau_D$ and $\tau_E$, where the former conveys information on the space exploration and the latter on the reproduction mechanism. Moreover, the pair correlation function will depend on whether population control is applied or not. In the following, we will denote by $h_f({{\mathbf x}_i},{{\mathbf x}_j},t)$ the pair correlation function for the case without population control, and $h_c({{\mathbf x}_i},{{\mathbf x}_j},t)$ for the case with population control.

\begin{figure}
  \includegraphics[width=\textwidth]{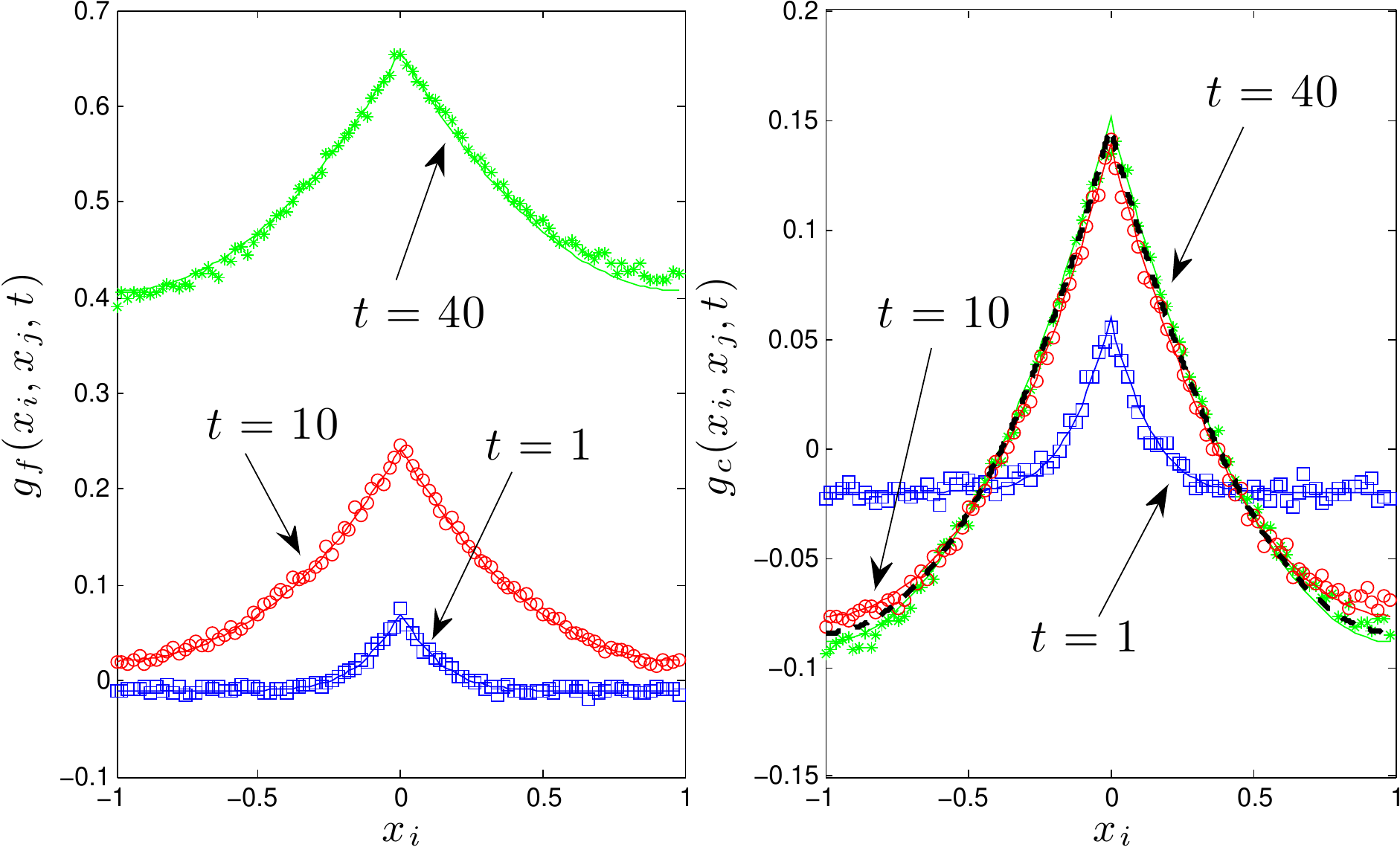}
\caption{The normalized and centered pair correlation function $g_{f,c}(x_i,x_j,t)$ for an initial collection of $N=10^2$ branching Brownian motions with diffusion coefficient $D=10^{-2}$ and birth-death rate $\lambda=1$ in a one-dimensional box of half-size $L=1$. We take $x_j=0$ and plot $g_{f,c}(x_i,x_j,t)$ with respect to $x_i$ at successive times $t=1$ (blue squares), $t=10$ (red circles) and $t=40$ (green stars). Symbols correspond to Monte Carlo simulations with $10^5$ ensembles, solid lines to exact solutions (Eqs.~\ref{free_corr_function} and~\ref{ref_correlation_function}, respectively). Left. For the case of a free system, $g_{f}(x_i,x_j,t)$ initially develops a peak at $x_i=x_j=0$, which is the signature of particles undergoing spatial clustering. At later times, $g_{f}(x_i,x_j,t)$ takes an asymptotic spatial shape, and is translated upwards by a spatially uniform term growing linearly in time. Right. For the case of a system with population control, $g_{c}(x_i,x_j,t)$ initially develops again a peak at $x_i=x_j=0$. Because of particle number conservation, an increased correlation about $x_i=0$ implies negative correlations close to $x_i=\pm L$. At later times, $g_{c}(x_i,x_j,t)$ converges to an asymptotic spatial shape $g^\infty_{c}(x_i,x_j)$ (Eq.~\ref{asygcontrol}), displayed as a black dashed curve.}
\label{fig4}
\end{figure}

For the simple reactor model detailed above, the pair correlation function $h_f({{\mathbf x}_i},{{\mathbf x}_j},t)$ for the free critical system has been derived in a previous work~\cite{zoia_pre_clustering} and is briefly recalled in Eq.~\ref{free_corr_function} in Section~\ref{calculations}. In this work we have explicitly derived the expression of $h_c({{\mathbf x}_i},{{\mathbf x}_j},t)$ for the case of a constant population in a confined geometry: the resulting formula is given in Eq.~\ref{ref_correlation_function}, and the derivation is provided in Section~\ref{calculations}. Once $h_{f,c}({{\mathbf x}_i},{{\mathbf x}_j},t)$ has been determined, it is customary to introduce the normalized and centered pair correlation function, in the form
\begin{equation}
g_{f,c}({{\mathbf x}_i},{{\mathbf x}_j},t) = \frac{ h_{f,c}({{\mathbf x}_i},{{\mathbf x}_j},t) - \langle n({{\mathbf x}_i},t) \rangle \langle n({{\mathbf x}_j},t) \rangle}{\langle n({{\mathbf x}_i},t) \rangle \langle n({{\mathbf x}_j},t) \rangle},
\end{equation}
which allows more easily comparing the amplitude of the typical spatial fluctuations to the average particle density.

\begin{figure}
  \includegraphics[width=0.7\textwidth]{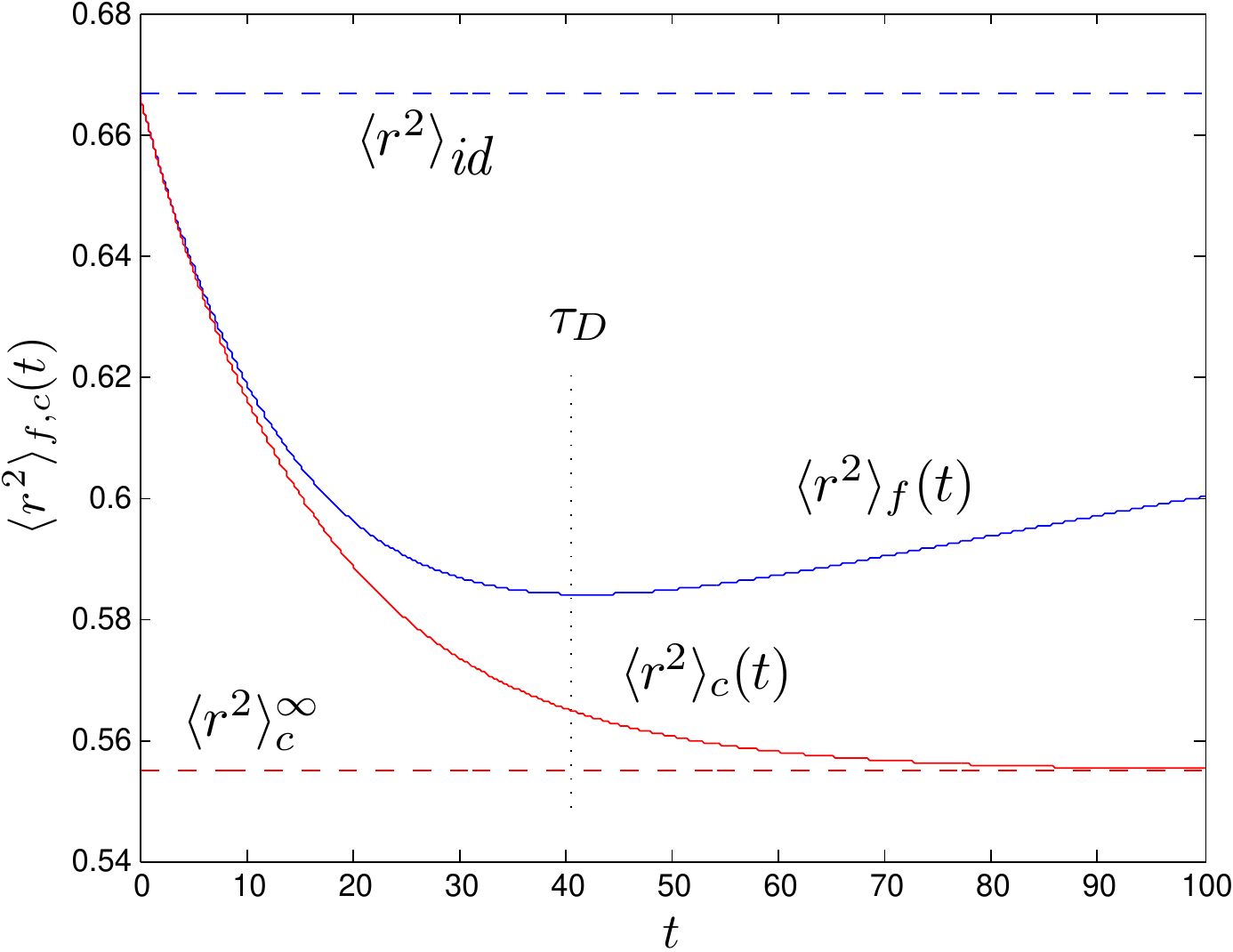}
\caption{The average square distance between particles $\langle r^2 \rangle_{f,c} (t)$ for the one-dimensional model with $N=10^2$ initial neutrons, $\lambda=1$, $D=10^{-2}$ and $L=1$. The blue solid curve corresponds to the free case: at long times, $\langle r^2 \rangle_{f} (t)$ asymptotically converges to the ideal average square distance $\langle r^2 \rangle_{ \textit{id}} = (2/3) L^2$ for a spatially uniform population, which is displayed as a blue dashed line. The red solid line corresponds to the case of population control: at long times, $\langle r^2 \rangle_{c} (t)$ asymptotically converges to the value $\langle r^2 \rangle_{c}^\infty$ given in Eq.~\ref{asy_x2_control}, which is displayed as a red dashed line.}
\label{fig5}
\end{figure}

For the purpose of physical analysis and illustration, let us consider an initial collection of $N=10^2$ neutrons subject to diffusion, reproduction and death in a one-dimensional bounded box $[-L,L]$, with $L=1$ (thus, $V=2$). To fix the ideas, we will set a diffusion coefficient $D=10^{-2}$ and a birth-death rate $\lambda= 1$. For this system, the mixing time reads $\tau_D \simeq 40.5$ and the extinction time reads $\tau_E \simeq 10^2$ (hence $\tau_D < \tau_E$; see Appendix~\ref{one_dimensional}). The initial condition for the neutron population is a uniform spatial distribution on $[-L,L]$. In Fig.~\ref{fig4}, we display the behaviour of the normalized and centered pair correlation functions $g_{f}(x_i,x_j,t)$ (left) and $g_{c}(x_i,x_j,t)$ (right) at successive times (their explicit expressions are reported in~\ref{one_dimensional}). We set $x_j=0$ and plot $g_{f,c}(x_i,x_j,t)$ as a function of $-L \le x_i \le L$. Solid curves represent the exact results given in Eqs.~\ref{free_corr_function} and~\ref{ref_correlation_function}, respectively, at three increasing times $t=1$, $t=10$ and $t=40$. Symbols represent Monte Carlo simulations performed with $10^5$ ensembles of $10^2$ neutrons.

For the free system, the pair correlation function $g_f(x_i,x_j,t)$ has three distinct regimes. Immediately after the initial time, $g_f(x_i,x_j,t)$ displays a peak at short distances $x_i \simeq x_j$, which mirrors the effects of local fluctuations responsible for spatial clustering. The amplitude of the peak is proportional to the ratio $\tau_D/\tau_E \propto \lambda L^2 / (N D)$ (see Eq.~\ref{gf_tau}), which precisely reflects the competition between migration and reproduction: the amplitude is larger for larger $D$ and smaller $\lambda$ (for fixed $L$ and $N$), and vice-versa. The width of the peak, which is related to the correlation length of the system, is governed by diffusion, and is a growing function of $D$. For the limit case of non-diffusing particles ($D \to 0$), $g_f(x_i,x_j,t)$ would display a delta-like behaviour at $x_i = x_j$, as expected on physical grounds: for long times, all the descendant particles have died, except for a few point-like clusters composed of a very large number of individuals. For times shorter than the mixing time $\tau_D$, the amplitude of the peak grows due to births and deaths dominating over diffusion, whereas its width increases due to diffusion. When $t \ge \tau_D$, the particles have explored the entire volume, and the tent-like shape of $g_f(x_i,x_j,t)$ freezes into its asymptotic behaviour (see Eq.~\ref{1d_frozen}).

The total number of neutrons in the reactor also undergoes global fluctuations due to the absence of population control and to $N$ being finite. These global fluctuations progressively lift upwards the shape of $g_f(x_i,x_j,t)$ by a spatially flat term that diverges linearly in time as $\sim \lambda \nu_2 t/ N$ (see Eq.~\ref{gf_tau}). Finally, for times larger than the extinction time $\tau_E$, $g_f(x_i,x_j,t) \ge 1$ (see Eq.~\ref{ext1d}). This physically means that, no matter how dense the system is at time $t=0$, global spatial fluctuations affect the whole volume with uniform (and increasing) intensity, and the neutrons are eventually doomed to extinction within a time $t \simeq N/ (\lambda \nu_2)$ in the absence of population control (see Fig.~\ref{fig1} b).

The average square distance between particles for the free system is displayed in Fig.~\ref{fig5}: at time $t=0$, the population is uniformly distributed and $\langle r^2 \rangle_{f}(0) = \langle r^2 \rangle_{id} = (2/3) L^2$. Immediately afterwards, $\langle r^2 \rangle_{f}(t)$ starts to decrease due to spatial clustering. For times longer than $\tau_D$, global fluctuations dominate, and correlations range over the whole box. Then, $\langle r^2 \rangle_{f}(t)$ increases and asymptotically saturates again to the ideal average square distance: this can be understood by observing that $h_f(x_i,x_j,t)$ becomes spatially flat for $t \gg \tau_E$.

\begin{figure}
  \includegraphics[width=8cm]{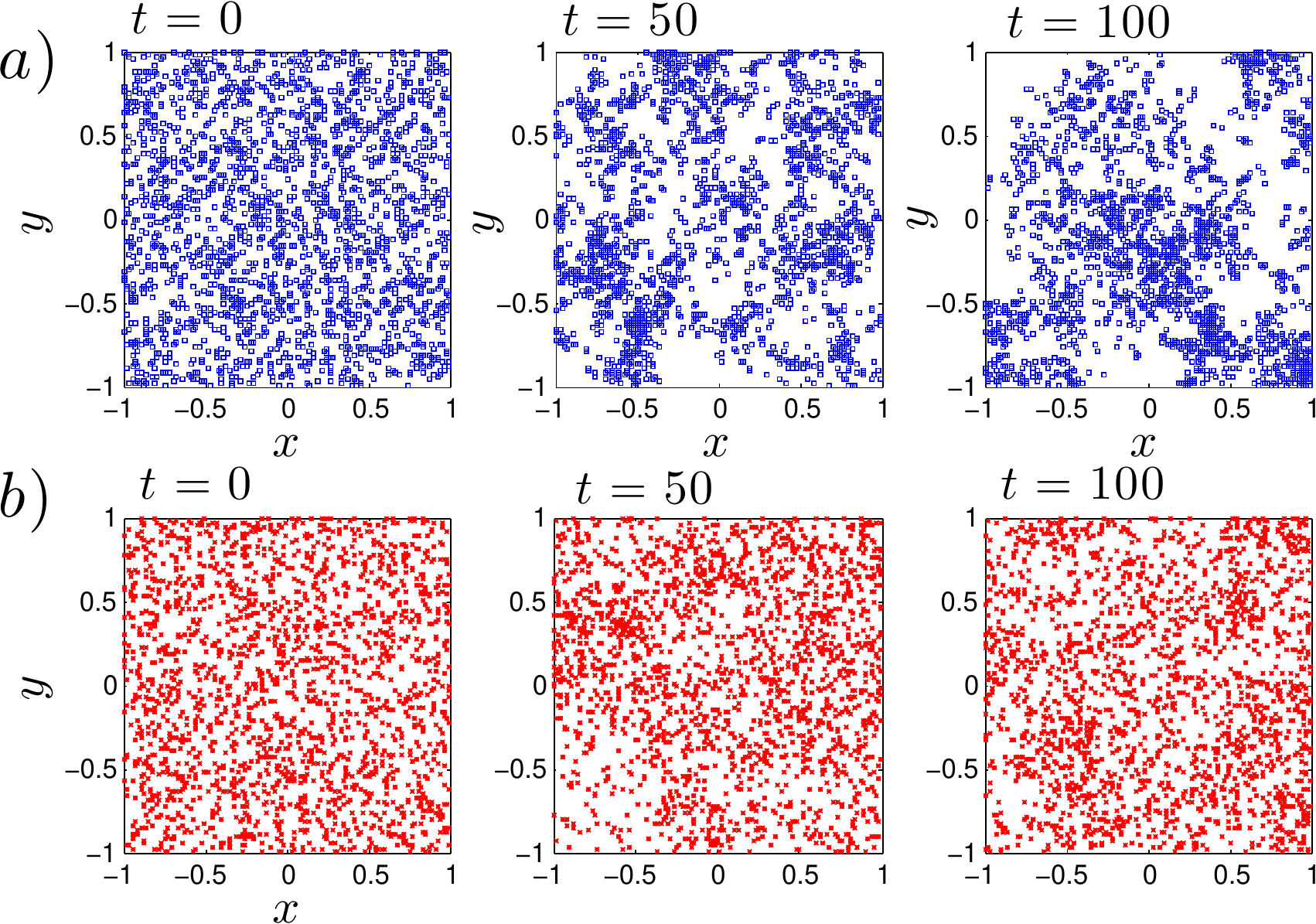}
\caption{Monte Carlo simulation of the evolution of a collection of branching Brownian motions in a two-dimensional box, subject to population control. The particles are prepared at $t=0$ on a uniform spatial distribution. In case $a)$, the ratio between the migration area and the the specific square separation distance between neutrons is small, namely, ${\cal A} / (\langle r^2 \rangle_{id}/N) \simeq 1.5$, and clustering phenomena dominate over diffusion (however, since the total particle number is preserved, the population can not go to extinction). In case $b)$, the ratio between the migration area and the the specific square separation distance between neutrons is large, namely, ${\cal A} / (\langle r^2 \rangle_{id}/N) \simeq 15$, and spatial fluctuations are much milder.}
\label{fig6}
\end{figure}

In the case of the system with population control, the pair correlation function $g_c(x_i,x_j,t)$ has two distinct regimes. Immediately after the initial time, spatial clustering effects are again reflected in a peak at short distances $x_i \simeq x_j$ for $g_c(x_i,x_j,t)$. The amplitude and the width of the peak have the same behaviour as for the free system detailed above. However, due to the conservation of the number of particles, the positive correlations at the center of the box imply now negative correlations close to the boundaries $x_i =\pm L$. For times shorter than the mixing time $\tau_D$, the amplitude of the peak grows and its width increases as in the previous case. Global spatial fluctuations are intrinsically suppressed by $N$ being fixed due to population control. For times larger than $\tau_D$, $g_c(x_i,x_j,t)$ converges to an asymptotic spatial shape $g^\infty_c(x_i,x_j)$ (see Eq.~\ref{asygcontrol}). In this regime, the amplitude of the pair correlation function is bounded by (see Eq.~\ref{boundg})
\begin{equation}
|g_c(x_i,x_j,t)| \le \frac{\langle r^2 \rangle_{id}}{N {\cal A}},
\end{equation}
where ${\cal A} = D/\lambda$ is the characteristic migration area of the particles, i.e., the square distance explored by diffusion during a generation, and $\langle r^2 \rangle_{id}/N$ is the specific square separation distance between neutrons corresponding to a uniform spatial distribution within the finite box~\cite{young}. In order for the fluctuations to be small and prevent the emergence of spatial clustering, we must therefore have ${\cal A} \gg \langle r^2 \rangle_{id}/N$, which occurs when the typical separation between particles is thoroughly explored within a single generation (see Fig.~\ref{fig6} for a numerical illustration). Observe that the equilibrium condition for a reactor to be operated at the critical point does not depend on the total neutron population $N$: therefore, in a system with population control, spatial clustering can be quenched by simply imposing that $N$ is sufficiently large, namely, $N \gg \langle r^2 \rangle_{id}/{\cal A}$.

The average square distance between particles for the system with population control is displayed in Fig.~\ref{fig5}: at time $t=0$, the population is uniformly distributed and we recover $\langle r^2 \rangle_{c}(0) = \langle r^2 \rangle_{id} = (2/3) L^2$. Immediately afterwards, $\langle r^2 \rangle_{c}(t)$ starts to decrease due to the competition between diffusion and birth-death. For times longer than $\tau_D$, $\langle r^2 \rangle_{c}(t)$ converges to the asymptotic value
\begin{equation}
\langle r^2 \rangle_{c}^\infty = \lim_{t \to \infty} \langle r^2 \rangle_{c}(t) = 4\frac{D}{\lambda_p} \left[1-\sqrt{\frac{2D}{\lambda_p L^2}} \tanh \left( \sqrt{\frac{\lambda_p L^2}{2D}} \right) \right],
\label{asy_x2_control}
\end{equation}
which generalizes to confined geometries the findings for unbounded domains derived in~\cite{meyer} (see~\ref{one_dimensional}).

\section{The pair correlation function}
\label{calculations}

The function $h_f({{\mathbf x}_i},{{\mathbf x}_j},t)$ for an exactly critical free system has been derived in~\cite{zoia_pre_clustering}, and can be written as $h_f = h_{f}^{(1)} + h_{f}^{(2)}$, where $h^{(1)}_f({{\mathbf x}_i},{{\mathbf x}_j},t) = h_{\textit{i}}({{\mathbf x}_i},{{\mathbf x}_j},t)$ is the contribution from uncorrelated trajectories, and
\begin{eqnarray}
h^{(2)}_f({{\mathbf x}_i},{{\mathbf x}_j},t) = \nonumber \\
\lambda\nu_2 N \int_0^{t} dt' \int_V d{\mathbf x}' {\cal G}({\mathbf x}_i, {\mathbf x}' ,t-t'){\cal G}({\mathbf x}_j, {\mathbf x}',t-t') \rho({\mathbf x}',t')
\end{eqnarray}
is the contribution of the trajectories leading from the final positions at ${\mathbf x}_i$ and ${\mathbf x}_j$ at time $t$ to the fission point ${\mathbf x}'$ at time $t'$. The coefficient $\nu_2=\sum_m m(m-1)\,q_m$ is the mean number of pairs created at each collision~\cite{bell_nuc}. The term $\lambda \nu_2 d t\, N\rho({\mathbf x},t) d{\mathbf x}$ therefore represents the average number of ordered pairs created about ${\mathbf x}$ in the time interval $(t, t+dt)$, which thus contribute to the correlation function about ${\mathbf x}_i$ and ${\mathbf x}_j$ at time $t$. Imposing a uniform spatial distribution $Q=1/V$ finally yields
\begin{equation}
h_f({{\mathbf x}_i},{{\mathbf x}_j},t) = \frac{N\left(N-1 \right)}{V^2} + \lambda\nu_2 N \int_0^{t} dt' {\cal G}({\mathbf x}_i, {\mathbf x}_j,2t').
\label{free_corr_function}
\end{equation}
The integral of the propagator appearing in Eq.~\ref{free_corr_function} is unbounded, so that at long times the amplitude of the correlations is expected to diverge.

The pair correlation function $h_c$ can be computed by closely following the arguments developed in~\cite{meyer}. Actually, the reactor model described above is basically identical to that proposed in~\cite{meyer}, but for boundary conditions (neutrons evolve in a confined geometry, whereas in~\cite{meyer} the viable space was unbounded) and initial conditions (in~\cite{meyer}, all the particles were located at the same point at $t=0$, whereas here the spatial distribution $\rho$ of the individuals at $t=0$ is arbitrary). Let us choose a pair of (distinct) neutrons located at ${\mathbf x}_i$ and ${\mathbf x}_j$ at time $t$. These neutrons may, or may not, have had a common ancestor (from a branching event) at a previous time $0<t'<t$. Because of particle number conservation, the fraction of new particle pairs created in the time interval $(t', t'+dt)$ is $\lambda_p dt = \lambda \nu_2 dt / (N-1)$, obtained as the ratio of the new particle pairs created in the time interval, i.e., $\lambda \nu_2 N d t /2$, to the total number of pairs $N(N-1)/2$~\cite{meyer}. The probability for a chosen pair of particles at time $t$ not to have had a common ancestor is $U(t)=e^{-\lambda_p t}$, so that the probability density for the ancestor to occur at time $t'$ for a particle pair observed at $t$ is
\begin{equation}
\psi_t(t') =\lambda_p \frac{U(t)}{U( t' )} =  \lambda_p e^{-\lambda_p (t-t')}.
\end{equation}
The functional form of $\psi_t(t')$ differs from that in~\cite{meyer}, since the initial conditions are different. The function $h_c({{\mathbf x}_i},{{\mathbf x}_j},t)$ can be again written as $h_c = h_{c}^{(1)} + h_{c}^{(2)}$, where $h_{c}^{(1)}({{\mathbf x}_i},{{\mathbf x}_j},t) = U(t)\, h_{\textit{i}}({{\mathbf x}_i},{{\mathbf x}_j},t)$ is the contribution of neutrons having evolved freely with no common ancestors. The correlated contribution reads
\begin{eqnarray}
h_{c}^{(2)}({{\mathbf x}_i},{{\mathbf x}_j},t) = \nonumber \\
N(N-1) \int_0^{t} \hspace{-1mm}  dt'\hspace{-1mm}  \int_V \hspace{-1mm}  d{\mathbf x}' {\cal G}({\mathbf x}_i, {\mathbf x}',t-t') {\cal G}({\mathbf x}_j, {\mathbf x}',t-t') \psi_t(t')  \rho({\mathbf x}',t').
\end{eqnarray}
where $N(N-1) \psi_t(t') d t'$ is the number of ordered particle pairs at time $t$ that have a common ancestor in the time interval $(t',t'+dt')$. The pair correlation function finally yields
\begin{equation}
h_c({{\mathbf x}_i},{{\mathbf x}_j},t) = \frac{N\left(N-1 \right)}{V^2}e^{-\lambda_p t}+\lambda\nu_2 \frac{N}{V} \int_0^{t}dt' e^{-\lambda_p t'} {\cal G}({\mathbf x}_i, {\mathbf x}_j, 2t')
\label{ref_correlation_function}
\end{equation}
when imposing $Q=1/V$. The integral of the propagator appearing in Eq.~\ref{ref_correlation_function} is bounded thanks to the exponential term, and at long times the correlation function converges to an asymptotic shape.

The specific shape of the correlation functions depends of the propagator, which can be computed once the dimension, geometry and boundary conditions of the problems have been assigned. In this respect, Eq.~\ref{ref_correlation_function} generalizes the result by~\cite{meyer} in that it allows for arbitrary geometries and initial conditions. The average square distance can be then obtained by direct integration by following Eq.~\ref{x2_definition}. In the Appendix, we develop the calculations for one-dimensional domains, which can be easily generalized to arbitrary dimension.

\section{Conclusions}
\label{conclusions}

In this work, we have analysed the statistical behaviour of fission chains in a prototype model of nuclear reactor. The neutron population evolving within a confined domain may display spatial clustering effects, due to the competition between births, deaths and diffusion. In particular, in the absence of population control, a reactor operated in stationary conditions at the critical point (corresponding to an exact equilibrium between the birth and death rates) would in principle lead to an almost sure extinction of the entire neutron population: this phenomenon goes under the name of critical catastrophe. Actually, such extinction is not observed in practice, because nuclear reactors are operated under strict population control policies (e.g., human intervention via control elements) that act a regularizing mechanisms at the global scale.

A key ingredient for our analysis of the space-time behaviour of the fluctuations is the pair correlation function, which is proportional to the joint probability density of finding particle pairs at two spatial sites at a given time. We have in particular explicitly derived and compared the pair correlation function of a spatially confined neutron population that is left free to evolve to that of a neutron population that is kept under control by demanding that the total number of individuals is constant at any time. We have shown that in the former case the pair correlation function at the critical point diverges linearly in time, whereas in the latter converges to an asymptotic spatial shape. For a free system, the ultimate fate is therefore extinction, no matter how dense the neutron population is at the initial time. For the system with population control, spatial fluctuations can be tamed by acting on the total number of neutrons that are present in the reactor at equilibrium: when the ratio of the specific square separation distance between neutrons corresponding to an ideal uniform spatial distribution and the neutron migration area (i.e., the typical square distance travelled by a particle by diffusion in the time span of a generation) is small, diffusion dominates over births and deaths, and the spatial clustering is quenched. This physically means that imposing a global feedback on the whole neutron population has a stabilizing effect also at the local scale of the spatial fluctuations.

The proposed reactor model retains the essential statistical aspects of actual systems, and as such it offers valuable insight on the behaviour of the spatial correlations. In our derivation, we have nonetheless introduced a number of simplifying hypotheses: for instance, we have assumed that the energy dependence of the physical parameters involved in neutron transport (such as the probability of fission and absorption, or the reaction rate) could be neglected; furthermore, we have modelled the boundaries of the reactor as reflecting, although a more realistic description would imply the possibility of particles leaking from the external surface. Finally, the effects of control rods in nuclear reactors are clearly localized in space, whereas in this work we have represented population control as a spatially homogeneous mechanism. Future work will be therefore aimed at investigating the impact of the introduced approximations.

\ack
The authors would like to thank Dr.~R.~Sanchez for stimulating discussions concerning the stochastic behaviour of the neutron population in nuclear reactors and for his critical reading of this manuscript.\\

\appendix
\section{Analysis of one-dimensional domains}
\label{one_dimensional}

Consider a finite box of half-size $L$, i.e., a segment $[-L,L]$ with $V=2L$. At the boundaries $x =\pm L$, we impose reflecting (Neumann) conditions. The propagator for this system reads~\cite{grebenkov}
\begin{equation}
{\cal G}(x, x_0, t) = \frac{1}{2L} + \frac{1}{L} \sum_{k=1}^\infty \varphi_k(x) \varphi_k(x_0) e^{- \alpha_k t },
\end{equation}
where we have set
\begin{equation}
\varphi_k(x) = \cos\left(\frac{k\pi (L-x)}{2L} \right)
\end{equation}
and
\begin{equation}
\alpha_k = \left(\frac{\pi}{2}\right)^2\frac{D}{L^2} k^2.
\end{equation}
We can identify the mixing time with $\tau_D =(2/\pi)^2 (L^2/D)$. If we choose the uniform spatial distribution $Q(x_0) = 1/2L$ at time $t=0$, the average density simply reads
\begin{equation}
\langle n(x_i, t) \rangle = \frac{N}{2L}.
\end{equation}
As for the pair correlation function, the case of the free system is obtained by resorting to Eq.~\ref{free_corr_function}, which yields
\begin{equation}
h_f(x_i,x_j,t) = \frac{N(N-1)}{(2L)^2} +\frac{\lambda\nu_2 N}{(2L)^2} \left[ t + \sum_{k=1}^\infty \varphi_k(x_i) \varphi_k(x_j) \frac{1-e^{-2\alpha_k t}}{\alpha_k} \right],
\label{1d_free_corr}
\end{equation}
For times $t \gg \tau_D$, the exponential term in Eq.~\ref{1d_free_corr} vanishes, and the spatial shape of $h_f(x_i,x_j,t)$ is frozen. In particular, the series appearing at the right-hand side is bounded, namely,
\begin{equation}
\sum_{k=1}^\infty \frac{\varphi_k(x_i) \varphi_k(x_j) }{\alpha_k} \le \frac{2}{3} \frac{L^2}{D}.
\label{1d_frozen}
\end{equation}
Then, for $N \gg 1$ the amplitude of the normalized and centered pair correlation function asymptotically grows as
\begin{equation}
g_f(x_i,x_j,t) = \frac{h_f(x_i,x_j,t) - \langle n(x_i, t) \rangle \langle n(x_j, t) \rangle}{\langle n(x_i, t) \rangle \langle n(x_j, t) \rangle} \simeq \frac{\lambda\nu_2 }{N} t.
\label{ext1d}
\end{equation}
Therefore, for times $t \gg N/(\lambda \nu_2)$, $g_f(x_i,x_j,t) \gg 1$, which allows identifying the extinction time $\tau_E = N/(\lambda \nu_2)$. Observe that, for $N \gg 1$, $g_f(x_i,x_j,t)$ can be expressed in terms of the two characteristic time scales, namely,
\begin{equation}
g_f(x_i,x_j,t) =  \frac{t}{\tau_E} + \frac{\tau_D}{\tau_E} \sum_{k=1}^\infty \varphi_k(x_i) \varphi_k(x_j) \frac{1-e^{-2 k^2 \frac{t}{\tau_D}}}{k^2}.
\label{gf_tau}
\end{equation}

The average square distance between particles
\begin{equation}
\langle r^2 \rangle_{f}(t) = \frac{\int dx_i \int dx_j (x_i - x_j)^2 h_{f}(x_i,x_j,t) }{\int dx_i \int dx_j h_{f}(x_i,x_j,t) }
\end{equation}
can be obtained by integration. At time $t=0$,
\begin{equation}
\langle r^2 \rangle_{f}(0) = \frac{2}{3}L^2 = \langle r^2 \rangle_\textit{id}.
\end{equation}
At times $t \gg \tau_E$, $h_{f}(x_i,x_j,t)$ becomes spatially flat, and $\langle r^2 \rangle_f(t)$ converges again to the ideal average square distance, namely, $\lim_{t \to \infty} \langle r^2 \rangle_{f}(t) = \langle r^2 \rangle_\textit{id} $.

For the case of population control, from Eq.~\ref{ref_correlation_function} we get
\begin{equation}
h_c(x_i,x_j,t) = \frac{N(N-1)}{(2L)^2} +\frac{\lambda\nu_2 N}{(2L)^2} \sum_{k=1}^\infty \varphi_k(x_i) \varphi_k(x_j) \frac{1-e^{-(2\alpha_k + \lambda_p)t}}{\alpha_k + \frac{\lambda_p}{2}},
\end{equation}
where $\lambda_p=\lambda/(N-1)$. Assuming that $\tau_E \gg \tau_D$, for times $t \gg \tau_D$ the series appearing at the right-hand side is bounded, namely,
\begin{equation}
\sum_{k=1}^\infty \frac{\varphi_k(x_i) \varphi_k(x_j)}{\alpha_k + \frac{\lambda_p}{2}}  \le \frac{\sqrt{\frac{2\lambda_p L^2}{D}} \coth \left(\sqrt{\frac{2\lambda_p L^2}{D}} \right) -1}{\lambda_p}.
\end{equation}
For $N \gg 1$ the normalized and centered pair correlation function at long times converges to
\begin{equation}
g_c^\infty(x_i,x_j) =\frac{\tau_D}{\tau_E} \sum_{k=1}^\infty \frac{\varphi_k(x_i) \varphi_k(x_j)}{k^2 + \frac{\tau_D}{2\tau_E}}.
\label{asygcontrol}
\end{equation}
In particular, its amplitude is asymptotically bounded by
\begin{equation}
|g_c(x_i,x_j,t)| \le \frac{\lambda\nu_2 }{N} \frac{2}{3} \frac{L^2}{D},
\label{boundg}
\end{equation}
where the absolute value is taken because $g_c(x_i,x_j,t)$ can be negative.

As for the average square distance, at time $t=0$ we have again $\langle r^2 \rangle_{c}(0) = \langle r^2 \rangle_\textit{id} $, as expected. The asymptotic behaviour of $\langle r^2 \rangle_{c}(t)$ at times $t \gg \tau_D$ can be computed exactly, and reads
\begin{equation}
\langle r^2 \rangle_{c}^\infty = \lim_{t \to \infty} \langle r^2 \rangle_{c}(t) = 4\frac{D}{\lambda_p} \left[1-\sqrt{\frac{2D}{\lambda_p L^2}} \tanh \left( \sqrt{\frac{\lambda_p L^2}{2D}} \right) \right].
\end{equation}
In the limit of extremely large populations, we have $\lim_{N \to \infty} \langle r^2 \rangle_{c} = (2/3) L^2$ and we recover the ideal case.\\


\end{document}